\documentclass[
aps,
prl,
reprint,
superscriptaddress,
floatfix,
nofootinbib
]{revtex4-2}

\usepackage[T1]{fontenc}
\usepackage[utf8]{inputenc}
\usepackage{graphicx}
\usepackage{amsmath}
\usepackage{amssymb}
\usepackage{bm}
\usepackage{booktabs}
\usepackage{xcolor}
\usepackage{hyperref}

\begin{document}

\title{From Superfluid Coherence to High-$K$ Fragment Channels at Nuclear
Scission}


\author{M.~Kowal}
\email{m.kowal@ncbj.gov.pl}
\affiliation{National Centre for Nuclear Research, Pasteura 7, 02-093 Warsaw, Poland}

\author{A.~Augustyn}
\affiliation{National Centre for Nuclear Research, Pasteura 7, 02-093 Warsaw, Poland}

\author{T.~Cap}
\affiliation{National Centre for Nuclear Research, Pasteura 7, 02-093 Warsaw, Poland}

\author{K.~Pomorski}
\affiliation{National Centre for Nuclear Research, Pasteura 7, 02-093 Warsaw, Poland}

\begin{abstract}
The origin of fission-fragment spin remains unresolved. 
For $^{236}$U fission, we combine deformation-dependent finite-temperature pairing
with a multidimensional scission-spectrum analysis. Near the pairing-quenching
boundary, pair breaking exposes near-Fermi high-$\Omega$ intruder orbitals and
opens high-$K$ channels. The resulting maximum spectroscopic capacities range
from a few to more than a dozen units of $\hbar$. Rapid neck rupture acts as
a non-adiabatic projection freeze-out, where pairing serves as a dynamical gate:
its critical attenuation does not generate angular momentum but allows existing
high-$K$ projections to be retained diabatically as local fragment-$K$ components
while total angular momentum remains conserved.
\end{abstract}

\maketitle

How paired fermionic matter accommodates angular momentum is a recurring
many-body question. In conventional superconductors, time-reversed states
with opposite momenta and spins form Cooper pairs~\cite{BCS1957}, whereas
Zeeman polarization can destabilize spin-singlet pairing at the Clogston
paramagnetic limit~\cite{Clogston1962}. In strongly interacting ultracold
Fermi gases, sufficiently large population imbalance suppresses
superfluidity~\cite{Zwierlein2006}, while rotation is carried by quantized
vortex lattices~\cite{Zwierlein2005}. In neutron-star crusts, vortex creep
and the catastrophic release of pinned vorticity were proposed as mechanisms
for pulsar timing irregularities and glitches~\cite{AndersonItoh1975}.
Together, these examples show that pairing constrains spin polarization,
whereas quantized vorticity provides a channel through which rotating
superfluids accommodate and transfer angular momentum.

Finite nuclei provide a mesoscopic realization of this physics. Pairing
couples time-reversed nucleons and reduces the moment of inertia below its
unpaired value~\cite{Belyaev1959,Migdal1959}. Increasing rotational frequency
weakens pairing through the Coriolis antipairing mechanism
~\cite{MottelsonValatin1960} and favors aligned high-$j$ two-quasiparticle
configurations~\cite{StephensSimon1972}. When such a configuration crosses
the paired ground-state rotational band, the moment of inertia changes
abruptly, producing the observed backbending~\cite{Johnson1971}. The internal
partition of angular momentum between collective rotation and quasiparticle
alignment is thereby reorganized, while total angular momentum remains
conserved.

Fission poses the inverse nonequilibrium problem: how does a low-spin paired
system divide into two fragments with substantial intrinsic spins? Decay
spectroscopy finds fragment-spin scales of several $\hbar$ and a pronounced mass
dependence~\cite{Wilhelmy1972,Wilson2021}, while partner-resolved measurements
show weak correlations between the spin magnitudes of the two
fragments~\cite{Wilson2021}. These weak correlations do not uniquely identify a
mechanism: neutron and $\gamma$ emission can substantially modify the primary
fragment spins after scission~\cite{Stetcu2021}, while pre-scission collective
dynamics can likewise produce weak correlations between the final fragment-spin
magnitudes~\cite{RandrupVogt2021,Scamps2023}.

\begin{figure*}[t]
\centering
\includegraphics[width=0.49\textwidth]{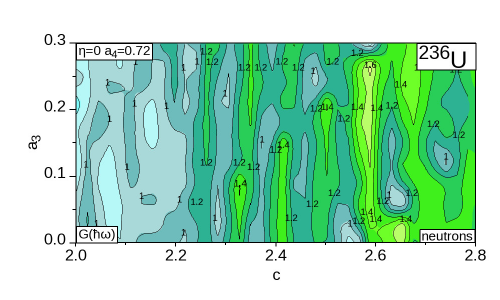}
\hfill
\includegraphics[width=0.49\textwidth]{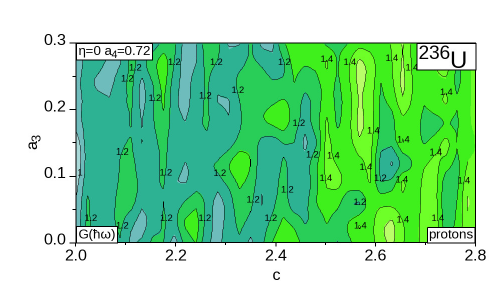}
\caption{Neutron $\Delta_{0n}(\mathbf q)$ (left) and proton
$\Delta_{0p}(\mathbf q)$ (right) zero-temperature pairing gaps entering
Eq.~(\ref{eq:gap}) in the $^{236}$U scission region, calculated with the
oscillator-scaled prescription
$G_\tau(\mathbf q)\propto\hbar\omega_0(\mathbf q)$. The maps are evaluated for
axial shapes ($\eta=0$) at the neck parameter $a_4=0.72$, with the elongation
$c$ and mass-asymmetry coordinate $a_3$ varied and the active higher-order
coordinates $a_5=a_6=0$. The coefficients $a_7$ and $a_8$ are fixed to zero
throughout the analysis. The corresponding heavy-fragment mass is approximately
$A_h\simeq A_0(1+0.9894a_3)/2$, where $A_0=236$.}
\label{fig:pairing_gaps_hw}
\end{figure*}

Microscopic calculations show that fragment deformation and shell structure
strongly influence the angular-momentum distributions of primary fission
fragments~\cite{Marevic2021,Bulgac2021}. However, the role of pairing in this
process remains an open question. Does pairing survive until rupture while
becoming sufficiently attenuated to permit the diabatic retention of
high-$\Omega$ quasiparticle configurations as mutually compensating fragment-$K$
components? Which near-Fermi branches then set the accessible projection scale?
To establish this pairing scale, we must reconcile two competing physical realities: the 
experimental footprint of odd--even staggering, which demands that pairing correlations 
survive until neck rupture, and the thermodynamic reality of thermal quenching. While multiple 
prescriptions for the coordinate dependence ($\mathbf q$) of the effective pairing strength $G_\tau(\mathbf q)$
are theoretically possible, the systematic analysis of these experimental and dissipative
constraints detailed in Ref.~\cite{SurvivalPairing} shows that constant-coupling and
surface-enhanced prescriptions lead to substantially different scission behavior. Here we adopt
the physically motivated oscillator-scaled prescription,
$G_\tau(\mathbf q)\propto\hbar\omega_0(\mathbf q)$. By tying the pairing strength to the local
major-shell spacing $\hbar\omega_0(\mathbf q)$, this prescription links the residual interaction to
the evolving mean-field geometry, maintaining physical consistency from compact ground states 
up to highly elongated scission configurations.
 Following the finite-temperature BCS pairing formalism established in Ref.~\cite{SurvivalPairing}, 
the thermal and  deformation evolution of the pairing gap is described by
\begin{equation}
\Delta_\tau(\mathbf q,T)=
\begin{cases}
\displaystyle
\Delta_{0\tau}(\mathbf q)\,
\tanh\!\left[
a\,\frac{T_{c\tau}(\mathbf q)-T}{\Delta_{0\tau}(\mathbf q)}
\right],
& T<T_{c\tau}(\mathbf q),\\[8pt]
0,
& T\geq T_{c\tau}(\mathbf q).
\end{cases}
\label{eq:gap}
\end{equation}
where $\tau=n,p$ labels the nucleon species and
$T_{c\tau}(\mathbf q)\simeq0.608\,\Delta_{0\tau}(\mathbf q)$ is the local critical temperature
above which the pairing gap disappears, and $a\simeq6.098$ is a dimensionless scaling 
parameter. At $T=T_{c\tau}(\mathbf q)$, the pairing field collapses and the corresponding pairing
contribution to the free energy vanishes. 
Thus deformation fixes the absolute pairing and
critical-temperature scales, whereas their dimensionless thermal
attenuation remains universal.
As an illustrative case, we consider $^{236}$U fission. The adopted prescription yields
characteristic zero-temperature scission gaps
$\Delta_{0\tau}(\mathbf q_{\rm sc})\simeq1.1\text{--}1.4$~MeV and critical temperatures
$T_{c\tau}(\mathbf q_{\rm sc})\simeq0.67\text{--}0.85$~MeV, where
$\mathbf q_{\rm sc}$ denotes a configuration on the scission hypersurface.
Since this critical temperature scale directly overlaps with the typical scission 
temperature $T_{\mathrm{sc}}\simeq0.7$ MeV extracted from Langevin trajectories~\cite{SurvivalPairing},
the scission ensemble is situated precisely across the pairing transition. This transitional 
regime preserves a fragile, residual paired component while allowing for sufficient 
pair-breaking and dissipation, bypassing the alternative scenarios detailed in 
Ref.~\cite{SurvivalPairing}. Mapped along the $^{236}\mathrm{U}$ scission line in 
Fig.~\ref{fig:pairing_gaps_hw}, these zero-temperature gaps $\Delta_{0\tau}(\mathbf q)$ 
and their finite-temperature counterparts [Eq.~(\ref{eq:gap})] establish the precise 
near-Fermi energetic window within which we can now investigate which specific 
high-$\Omega$ single-particle orbitals become accessible during rapid neck rupture.

The next question is which single-particle orbitals occupy this
pairing-controlled window. Spin--orbit coupling lowers high-$j$ intruder
orbitals toward the Fermi energy, while strong axial elongation makes the
fission axis the natural quantization axis and splits their multiplets into
Nilsson branches with different projections $\Omega$.\footnote{We use the standard Nilsson notation
$[N\,n_z\,\Lambda]\Omega$. Within this label, $N$ denotes the major
oscillator-shell quantum number, not the neutron number introduced above;
$n_z$ counts oscillator quanta along the symmetry axis; $\Lambda$ and
$\Sigma=\pm1/2$ are the orbital and spin projections on that axis; and
$\Omega=\Lambda+\Sigma$ is the total single-particle projection. For a
multiquasiparticle configuration,
$K=\left|\sum_i\Omega_i\right|$, where $i$ labels the quasiparticles. Two
quasiparticles generate the Gallagher--Moszkowski branches
$K=|\Omega_1\pm\Omega_2|$}
As the neck develops, these branches acquire different deformation slopes.
Some high-$\Omega$ components approach the Fermi surface and localize in the
nascent fragments, providing possible carriers of uncompensated axial
projection when time-reversed pairs are broken.

Their proximity to the Fermi surface follows from two effects. First, the
spin--orbit interaction lowers the spherical $j=l+1/2$ members of high-$l$
shells, bringing high-$j$ intruder orbitals from the next major shell into the
relevant energy region. Second, axial deformation splits each multiplet into
branches with $\Omega=\Lambda+\Sigma$, where $\Sigma=\pm1/2$. The
strong-coupling shift
$\Delta E_{\rm ls}\simeq-2\kappa\hbar\omega_0\Lambda\Sigma$, with $\kappa$
the spin--orbit strength, additionally lowers the branches with aligned orbital
and spin projections. The quadrupole field gives these branches different
deformation slopes and selects which ones reach the Fermi surface. High $K$
does not itself lower an orbital: spin--orbit coupling brings the intruder shell
down, and deformation selects its accessible high-$\Omega$ components.

Pairing normally cancels these projections. Time-reversed states with
$+\Omega$ and $-\Omega$ form $K=0$ pairs, so their separate contributions are
not expressed. Pair attenuation removes this constraint. If the resulting
quasiparticles localize in different prefragments, or their occupations freeze
into different diabatic configurations during rupture, nonzero local
fragment-$K$ components may emerge while the total projection remains
conserved.

All of these spectroscopic insights of 
high-$K$ projections hinge fundamentally on the physical conditions at the scission 
moment. Yet, defining and identifying scission remains a highly non-trivial and 
crucial challenge in fission theory. To address this, we generalize the concept of 
scission by treating it multidimensionally rather than as a single, unique line in 
collective space. Operationally, we characterize this scission boundary as a 
hypersurface---the ``scission wall''---defined by the minimum neck radius $r_{\rm neck}$. 
Its onset is identified by the first configurations with $r_{\rm neck}<1.5$~fm reached 
from the ground-state side. To systematically sample the nuclear shapes immediately 
preceding rupture, we also consider the deeper asymmetric region where 
$1.2~{\rm fm}<r_{\rm neck}<1.3$~fm and $f_h\equiv V_h/V_0=A_h/A_0>0.6$.
Crucially, this wall is mapped within the
five-dimensional axial deformation space introduced below rather than along a simplified,
one-dimensional path.

For each selected configuration on the scission hypersurface, we diagonalize the deformed 
Woods--Saxon single-particle Hamiltonian to resolve the near-Fermi neutron and proton 
spectra. This spectroscopic analysis is performed in the five-dimensional axial
Fourier-over-Spheroid space $\mathbf q=(c,a_3,a_4,a_5,a_6)$~\cite{Pomorski2023FoS}, with axial
symmetry imposed by setting $\eta = 0$ and with $a_7=a_8=0$ throughout. Within this
representation, $c$ governs elongation, $a_3$ regulates mass asymmetry, $a_4$ controls
neck development, and $a_5$ and $a_6$ describe higher-order axial surface distortions.
By utilizing
this expanded coordinate space, the local single-particle structure---and thus the 
possible $K$-carrying configurations---is determined by the detailed neck geometry 
rather than by simplified macroscopic coordinates alone. We employ the universal 
Woods--Saxon parametrization~\cite{Cwiok1987WoodsSaxon}, which is our standard 
spectroscopic input in systematic studies of high-$K$ configurations in the heaviest 
nuclei~\cite{Jachimowicz2015HighK,Jachimowicz2023MdRg,Jachimowicz2025FmCn,Brodzinski2017SHN,Jachimowicz2018Hindered}.

\begin{figure}[t]
\centering
\includegraphics[width=\columnwidth]{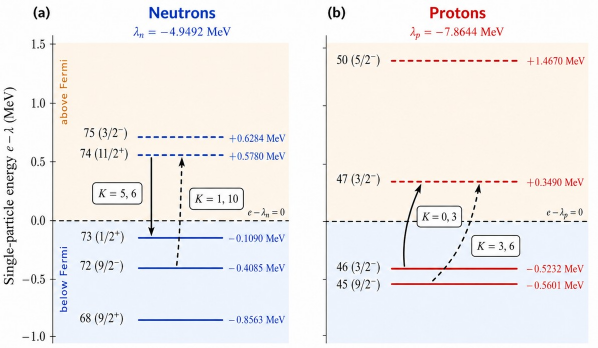}
\caption{Near-Fermi neutron (left) and proton (right) levels relative to
$\lambda_n=-4.9492$~MeV and $\lambda_p=-7.8644$~MeV. Blue/solid and
orange/dashed horizontal lines denote occupied and unoccupied states. Solid
arrows show the nearest-level pair-breaking branches; dashed arrows show
representative branches admitted by the extended prescription. Labels give the
Gallagher--Moszkowski values of $K$. The representative scission configuration is $c=2.200$,
$a_3=0.200$, $a_4=0.700$, and $a_6=-0.025$, with the remaining active coordinate
$a_5=0$; as throughout the analysis, $a_7=a_8=0$.}
\label{fig:single_particle_scheme}
\end{figure}

We first test the mechanism at one representative point on the scission line.
Figure~\ref{fig:single_particle_scheme} shows the near-Fermi levels and
candidate pair-breaking branches. Energies $e$ are measured from the neutron
and proton Fermi surfaces, $\lambda_n$ and $\lambda_p$; negative and positive
values identify occupied hole and unoccupied particle states, respectively.
The arrows denote spectroscopically allowed branches, not calculated transition
amplitudes or populations.

We compare two state-selection rules. The strict prescription retains only the
occupied and unoccupied orbitals closest to the Fermi surface in each nucleon
sector. The extended prescription also admits nearby high-$\Omega$ orbitals
within an energy window determined by the local excitation of the system. 
Rather than choosing an arbitrary cutoff, we use the pairing scale at the representative
scission temperature $T_{\mathrm{sc}}=0.70$~MeV. In the Fermi-gas estimate
$E_{\rm int}=a_{\rm ld}T_{\rm sc}^{2}$, with
$a_{\rm ld}=A/8.5~{\rm MeV}^{-1}$ and $A=236$, this temperature corresponds to
$E_{\rm int}\simeq13.6$~MeV. Under the universal gap-attenuation relation
[Eq.~(\ref{eq:gap})], this temperature reduces a local zero-temperature gap of
$\Delta_0\simeq1.36$~MeV ($T_c\simeq0.83$~MeV) to the thermally attenuated scale
$\Delta=0.70$~MeV, which we adopt as a pairing-motivated selection window for
accessible quasiparticle states. Comparing these two prescriptions tests whether energy ordering
alone misses structurally protected high-$K$ channels.

Throughout, $\mathcal K_{\rm spec}^{\max}$ denotes the maximum spectroscopic
capacity of the locally available quasiparticle channels. For orbitals with
projections $\Omega_1$ and $\Omega_2$, the largest
Gallagher--Moszkowski branch is $\max(|\Omega_1-\Omega_2|,
\Omega_1+\Omega_2)$. When neutron and proton pairs are both broken, their
largest aligned contributions are added. This construction gives an upper
capacity, not a population-weighted observable. It is neither the conserved
projection of the complete system nor a predicted fragment spin. In an axially
symmetric evolution, nonzero local projections must occur in compensating
combinations so that the initial total projection is preserved.

Under the strict prescription, the neutron states closest to the Fermi surface
are $1/2^+$ at $e-\lambda_n=-0.109$~MeV and $11/2^+$ at $+0.578$~MeV. They
give $K=5,6$ and $K_n^{\max}=6$. The corresponding proton states are $3/2^-$
at $e-\lambda_p=-0.523$~MeV and $3/2^-$ at $+0.349$~MeV, giving $K=0,3$ and
$K_p^{\max}=3$. The most conservative aligned capacity of this shape is
$\mathcal K_{\rm spec}^{\max}\simeq6+3=9$.

The extended prescription also includes the occupied neutron $9/2^-$ orbital
at $e-\lambda_n=-0.409$~MeV. Coupled to the $11/2^+$ particle state, it gives
$K=1,10$ [dashed arrow in Fig.~\ref{fig:single_particle_scheme}(a)], raising
$K_n^{\max}$ to 10 and, while retaining the strict prescription for protons,
$\mathcal K_{\rm spec}^{\max}$ to approximately 13. The hole state is dominated
by the Nilsson components $[11\,7\,4]$, $[10\,6\,4]$, and $[12\,8\,4]$; the
particle state by $[8\,3\,5]$, $[9\,4\,5]$, and $[7\,2\,5]$. Their dominant
orbital projections, $\Lambda=4$ and 5, differ by only one unit of angular 
momentum. Due to the centrifugal barrier associated with high orbital projections,
their wavefunctions are strongly suppressed near the symmetry axis; in this region,
their radial dependence scales as $R(\rho)\propto\rho^{|\Lambda|}$. Consequently,
the associated quasiparticle densities are strongly suppressed in the neck region,
rendering them structurally resilient against the rapid, non-adiabatic geometry changes 
driving the final stages of rupture.\footnote{This similarity refers only to the axial suppression of the two
densities. The eigenstates remain orthogonal, no transition matrix element is
implied, and equal localization away from the axis is not established. The
states also belong to different $\Omega$ blocks and are not directly mixed by
a strictly axial perturbation. Their high-$\Lambda$ content supports reduced
coupling to the neck, not a population estimate for the $K=10$ branch.}

The energetically closest neutron configuration is therefore not necessarily
the most resistant to rupture. The nearest occupied $1/2^+$ state has dominant
$\Lambda=0$ character and can penetrate the neck, whereas the
$9/2^-\otimes11/2^+$ branch is built from $\Lambda=4$ and 5 components. Energy
remains essential, but energy-only selection can underestimate nearby high-$K$
configurations whose densities avoid the neck. This structural argument does
not prove that the $K=10$ branch is more strongly populated.

The proton spectrum contains an analogous near-degeneracy. The occupied
$9/2^-$ orbital at $e-\lambda_p=-0.560$~MeV lies only $0.037$~MeV farther from
the Fermi surface than the occupied $3/2^-$ orbital at $-0.523$~MeV. Coupling
it to the unoccupied $3/2^-$ level gives $K=3,6$ [dashed arrow in
Fig.~\ref{fig:single_particle_scheme}(b)] and raises $K_p^{\max}$ to 6. With the
neutron $9/2^-\otimes11/2^+$ branch,
$\mathcal K_{\rm spec}^{\max}\simeq16$. This is a mixed four-quasiparticle
channel built from neutron and proton two-quasiparticle components. The sequence
$9\to13\to16$ is the point of the example: the nearest levels provide moderate
projection capacity, while near-degenerate high-$\Omega$ orbitals open the
high-$K$ sector without invoking remote excitations.

\begin{figure}[t]
\centering
\includegraphics[width=\columnwidth]{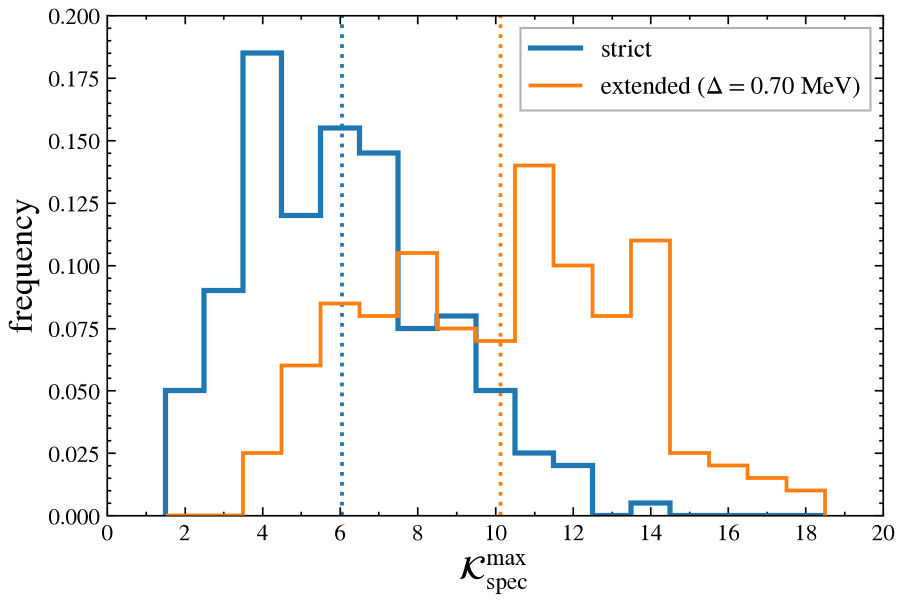}
\caption{Distribution of the maximum aligned pre-scission spectroscopic
capacity $\mathcal K_{\rm spec}^{\max}$ across 200 sampled configurations.
Blue histograms show the strict nearest-level prescription; orange histograms
include high-$\Omega$ orbitals within the pairing-attenuated window ($\Delta = 0.70$~MeV) of the Fermi surface. Dotted
lines denote the sample means. Because the configurations are not weighted by
dynamical fluxes or transition probabilities, the histogram is not a predicted
fragment-spin distribution.}
\label{fig:k_histogram}
\end{figure}

To test whether this representative configuration captures the general behavior,
we repeat the extraction for 200 random configurations in the deeper asymmetric
region defined above, satisfying $1.2~{\rm fm}<r_{\rm neck}<1.3$~fm and $f_h>0.6$.
The sampled configurations are therefore not restricted to splits near $3:2$;
their mean heavy-fragment fraction is $\langle f_h\rangle\simeq0.70$.
For every shape, we construct the
neutron and proton Gallagher--Moszkowski branches and record
$\mathcal K_{\rm spec}^{\max}$. Figure~\ref{fig:k_histogram} shows the resulting
ensemble. The strict prescription gives
$\langle\mathcal K_{\rm spec}^{\max}\rangle\simeq6\hbar$. When high-$\Omega$
orbitals within the pairing-attenuated window ($\Delta = 0.70$~MeV) of the Fermi surface are included, the mean shifts to
approximately $10\hbar$ and a clear tail develops beyond $15\hbar$. High-$K$
orbitals need not be the closest states, but they occur within a narrow
near-Fermi window across the sampled scission region. Pair breaking can expose
these channels; their populations require a dynamical calculation.

Timing, not just energy, determines retention. Pairing facilitates adiabatic occupation
rearrangement across the evolving single-particle spectrum; its attenuation
makes the last stage partly diabatic. Time-dependent calculations give proton
and neutron neck-decay times of approximately $\tau_p=0.05$~zs and
$\tau_n=0.12$~zs, both shorter than the minimum interfragment communication
time of approximately $0.5$~zs~\cite{Abdurrahman2024,Ren2022,Li2023}.
Occupations inherited from the connected system may therefore freeze before
relaxing to the equilibrium fragment configurations.

Returning to the representative neutron branch $9/2^-\otimes11/2^+$ introduced
in Fig.~\ref{fig:single_particle_scheme}(a),
$e_h-\lambda_n=-0.409$~MeV and $e_p-\lambda_n=0.578$~MeV. Using the
representative thermally attenuated gap $\Delta=0.70$~MeV and
$E_i=[(e_i-\lambda_n)^2+\Delta^2]^{1/2}$ gives $E_h=0.811$~MeV,
$E_p=0.908$~MeV, and $E_{2{\rm qp}}=1.719$~MeV. The inverse spectral scale is
$\tau_{\rm spec}=\hbar/E_{2{\rm qp}}=0.38$~zs, so
$\tau_p/\tau_{\rm spec}\simeq0.13$ and
$\tau_n/\tau_{\rm spec}\simeq0.31$. Rupture is sudden on the relevant
quasiparticle scale. This supports, but does not determine the probability of,
diabatic retention. The large-$|\Lambda|$ components of high-$\Omega$ orbitals
may be especially robust because their suppression near the symmetry axis
reduces coupling through the narrow neck.

Pairing does not store net angular momentum. It binds time-reversed states into
$K=0$ pairs and suppresses their separate projections. Pair breaking lifts
this constraint, allowing mutually compensated projections to localize as
nonzero fragment-$K$ components while total angular momentum remains
conserved~\cite{Scamps2026}. Pairing thus operates as a microscopic dynamical 
gate: while the fully paired superfluid phase systematically screens intrinsic spin 
projections, its critical suppression near rupture lifts this topological 
constraint, allowing the latent high-$K$ single-particle structure to manifest 
as primary fragment spin.

In summary, deformation places $^{236}$U scission near the pairing-quenching
boundary and simultaneously brings high-$\Omega$ intruder orbitals into the
near-Fermi window. Across the sampled scission configurations, these orbitals
increase the spectroscopic capacity and generate a high-$K$ tail absent from
the nearest-level prescription. Neck rupture is faster than the representative
quasiparticle response, allowing the exposed, globally compensated projections
to be retained diabatically as local fragment-$K$ components. The critical
attenuation and subsequent non-adiabatic collapse of nuclear superfluidity at 
rupture therefore establish a compelling microscopic mechanism for fragment-spin
formation. This process provides a general paradigm for how low-spin, 
strongly paired many-body systems partition into spinning, decoupled 
subsystems—a phenomenon with potential analogies in other mesoscopic superfluid 
and superconducting systems undergoing topological phase transitions.

If near-Fermi high-$\Omega$ orbitals are populated diabatically, the corresponding fragments should exhibit an enhanced high-spin tail, even if their mean spin changes only moderately. We can test this by comparing calculated high-$\Omega$ spectral regions with prompt-$\gamma$ spin reconstructions and isomeric-yield ratios. Odd--even staggering independently tests pairing survival in the same regions. Our calculation identifies the available high-$K$ configurations, not their populations; quantifying the enhancement requires dynamical pair-breaking and occupation-transfer calculations.

\begin{acknowledgments}
This research was funded in part by the National Science Centre, Poland, under
Project No.~2023/49/B/ST2/01294. M.~Kowal and T.~Cap were also partially
supported by the Polish--French cooperation COPIGAL.
\end{acknowledgments}

\bibliographystyle{apsrev4-2}
\bibliography{references}

\end{document}